\documentclass[12pt,preprint]{aastex}
\usepackage{graphicx}
\usepackage{ulem}

\begin{document}
\title{Multi-wavelength {\it Hubble Space Telescope} photometry of stellar populations in NGC\,288.
          \footnote{           Based on observations with  the
                               NASA/ESA {\it Hubble Space Telescope},
                               obtained at  the Space Telescope Science
                               Institute,  which is operated by AURA, Inc.,
                               under NASA contract NAS 5-26555.}}
\author{
 G.\ Piotto\altaffilmark{2,3},
 A.\ P. \,Milone\altaffilmark{4,5,6}, 
 A.\ F. \,Marino\altaffilmark{4},
 L.\ R. \,Bedin\altaffilmark{3},
 J.\ Anderson\altaffilmark{7},  
 H.\ Jerjen\altaffilmark{4},
 A.\ Bellini\altaffilmark{7}, 
 S.\ Cassisi\altaffilmark{8}
 }

\altaffiltext{2}{Dipartimento  di  Fisica e Astronomia, `Galileo Galilei' Universit\`a  di Padova, Vicolo dell'Osservatorio 3, Padova I-35122, Italy; giampaolo.piotto@unipd.it }

\altaffiltext{3}{INAF-Osservatorio Astronomico di Padova, Vicolo dell'Osservatorio 5, I-35122 Padua, Italy; luigi.bedin@oapd.inaf.it}

\altaffiltext{4}{Research School of Astronomy and Astrophysics, The Australian National University, Cotter Road, Weston, ACT, 2611, Australia
 milone@mso.anu.edu.au, amarino@mso.anu.edu.au, jerjen@mso.anu.edu.au}

\altaffiltext{5}{Instituto de Astrof\`\i sica de Canarias, E-38200 La Laguna, Tenerife, Canary Islands, Spain}

\altaffiltext{6}{Department of Astrophysics, University of La Laguna, E-38200 La Laguna, Tenerife, Canary Islands, Spain}

\altaffiltext{7}{Space Telescope Science Institute, 3800 San Martin Drive, Baltimore, MD 21218; jayander@stsci.edu, bellini@stsci.edu}

\altaffiltext{8}{INAF-Osservatorio Astronomico di Collurania, via Mentore Maggini, I-64100 Teramo, Italy; cassisi@oa-teramo.inaf.it} 

\begin{abstract}
We present new UV observations for NGC\,288, taken with the WFC3 detector
on board the {\it Hubble Space Telescope}, and combine them with existing
optical data from the archive to explore the multiple-population phenomenon
in this globular cluster (GC).  The WFC3's UV filters have demonstrated an uncanny 
ability to distinguish multiple populations along all photometric sequences
in GCs, thanks to their exquisite sensitivity to the 
atmospheric changes that are tell-tale signs of second-generation enrichment.
Optical filters, on the other hand, are more sensitive to stellar-structure 
changes related to helium enhancement.  By combining both UV and optical 
data we can measure helium variation.  We quantify
this enhancement for NGC\,288 and find that its variation is typical
of what we have come to expect in other clusters.
\end{abstract}


\shorttitle{Multiple populations in NGC\,288} 
\shortauthors{Piotto et al.} 

\section{Introduction}
\label{introduction}
Recent observations with the {\it Hubble Space Telescope} (HST) have shown
that color-magnitude diagrams (CMDs) of globular clusters (GCs) are very
different from our classical expectations of razor-thin sequences 
characteristic of single, old populations of stars.  In particular, 
{\it HST\,} near-UV data has shown that most, if not all, GCs host multiple 
stellar populations, as evidenced by two or more intertwined sequences in 
the CMD that we can trace from the main sequence (MS), through the 
sub-giant branch (SGB), up the red-giant branch (RGB) and even along the 
horizontal branch (HB).

>From studies of several clusters, we have found that the different sequences 
can vary their color separation or even invert their relative colors, 
depending on the photometric-band combinations.  These sequences correspond 
to stellar populations that have different abundances of light elements and 
helium.  A comparison of the photometry with synthetic spectra can provide 
unique opportunities to estimate the helium content among the stellar 
populations (e.g.\ Milone et al.\ 2012a), even at the level of faint 
MS stars, which are unreachable by spectroscopic investigations.

The cluster analysed in this paper, NGC\,288, is already known to host 
two populations of stars characterized by difference in light-element 
abundance (e.g.\ Shetrone \& Keane 2000, Kayser et al.\ 2008, Smith \& 
Langland-Shula 2009,  Carretta et al.\ 2009, Pancino et al.\ 2010).  The 
RGB of NGC\,288 is bimodal, when observed in appropriate ultraviolet filters,
and each RGB is populated by stars with different abundance of sodium and 
oxygen (Lee et al.\ 2009, Roh et al.\ 2011, Monelli et al.\ 2013). 

In this paper, we combine new {\it HST} observations with archival data
to investigate the evolutionary path of the multiple populations in NGC\,288 
along the MS, SGB, and RGB.  By exploring a wide wavelength region, ranging 
from the ultraviolet ($\sim$2750\AA) to the near infrared ($\sim$8140\AA), 
we will estimate the helium difference between the two main populations.

\section{Data and Data Reduction}
\label{sect:data}
To get the broadest possible perspective on NGC\,288's multiple 
populations, we consolidated photometry from a large number of
{\it HST} images taken with the Wide Field Channel of the Advanced 
Camera for Surveys (ACS/WFC) and the ultraviolet/visible channel 
of the Wide-Field Camera 3 (WFC3/UVIS).  Table~1 gives a list of 
the data sets we used.  Most {\it HST} data are from the archive, 
with the exception of the proprietary images from GO-12605 (PI:\ Piotto), 
which were specifically taken for this project and are crucial for 
its success.

Photometric and astrometric measurements of ACS/WFC exposures were 
obtained with the software program described by Anderson et al.\ (2008).  
This routine produces a catalog of stars over the field of view by 
analyzing an entire set of images simultaneously.  It measures stellar
fluxes independently in each exposure by means of a spatially variable 
point-spread-function model (see Anderson \& King 2006), along with a 
spatially-constant perturbation of the PSF to account for the 
effects of focus variations.  The photometry has been calibrated as 
in Bedin et al.\ (2005), using the encircled energy and zero points of 
Sirianni et al.\ (2005). 

The WFC3/UVIS images were reduced as described in Bellini et al.\ (2010), 
with\\ 
\textit{img2xym\_UVIS\_09$\times$10}, a software routine that is 
adapted from \textit{img2xym\_WFI} (Anderson et al.\ 2006).  Astrometry 
and photometry were corrected for pixel area and geometric distortion 
as in Bellini \& Bedin (2009), and Bellini, Anderson \& Bedin (2011).  
There are a few filters for which filter-specific distortion solutions are 
not yet available.  For these filters (F395N, F467M, and F547M), we applied
the solution for the closest available filter.  This introduces small 
(0.05 pixel) errors in astrometry and negligible errors in photometry.

Since the main results of this paper require high-precision 
photometry, we limited our analysis to the sub-sample of stars well measured.  The software routine provides several quality 
indexes that can be used as diagnostics of the reliability of 
photometric measurements: 
  i) the rms of the individual position measurements about their mean, 
     after they have been measured in different exposures and transformed 
     into a common reference frame ($rms_{\rm X}$ and $rms_{\rm Y}$ ), 
 ii) $o$, the ratio between the estimated flux of the star in a 0.5 arcsec 
     aperture and the flux from neighbor stars that has spilled over into
     the same aperture), and
iii) $q$, the residuals to the PSF fit for each star (see 
     Anderson et al.\ 2008 for details).  
To select the high-quality sub-sample of stars we followed the approach
described by Milone et al.\ (2009, Sect.~2.1).  Photometry has been 
corrected for differential reddening by means of a procedure that
has been adopted for several other projects and is described in detail 
in Milone et al.\ (2012b).  Briefly, we define the fiducial MS for the 
cluster and then identify for each star a set of neighbors and determine
from them their median offset relative to the fiducial sequence;
this systematic color and magnitude offset, measured along the reddening 
line, is our estimate of the local differential-reddening value.

\begin{table}[!htp]
\center
\scriptsize {
\begin{tabular}{cccccl}
\hline
\hline
 INSTR.\, &  DATE & N$\times$EXPTIME & FILTER  & PROGRAM & PI \\
\hline
ACS/WFC & Sep 20 2004 & 60s$+$680s & F435W & 10120 &  S.\,Anderson\\
ACS/WFC & Sep 20 2004 & 10s$+$75s$+$115s$+$120s& F625W & 10120 &  S.\,Anderson\\
ACS/WFC & Sep 20 2004 & 680s$+$1080s& F658N & 10120 &  S.\,Anderson\\
ACS/WFC & Jul 31 2006 & 10s$+$4$\times$130s & F606W & 10775 & A.\,Sarajedini\\
ACS/WFC & Jul 32 2006 & 10s$+$4$\times$150s & F814W & 10775 & A.\,Sarajedini\\
WFC3/UVIS & Nov 11 2011 & 2$\times$360s & F547M & 12193 &  J.\,W.\,Lee \\
WFC3/UVIS & Nov 11 2011 & 964s$+$1055s  & F467M & 12193 &  J.\,W.\,Lee \\
WFC3/UVIS & Nov 11 2011 & 1260s$+$1300s & F395N & 12193 &  J.\,W.\,Lee \\
WFC3/UVIS & Oct 25 and Nov 7 2012 & 4$\times$400s$+$2$\times$401s & F275W & 12605 & G.\,Piotto \\
WFC3/UVIS & Oct 25 and Nov 7 2012 & 4$\times$350s & F336W & 12605 & G.\,Piotto \\
WFC3/UVIS & Oct 25 and Nov 7 2012 & 4$\times$41s  & F438W & 12605 & G.\,Piotto \\
\hline
\hline
\end{tabular}
}
\label{tab:data}
\caption{List of the data sets used in this paper. }
\end{table}

\section{The color-magnitude diagram}
\label{sec:cmd}
\begin{figure}[htp!]
   \centering
   \epsscale{.75}
      \plotone{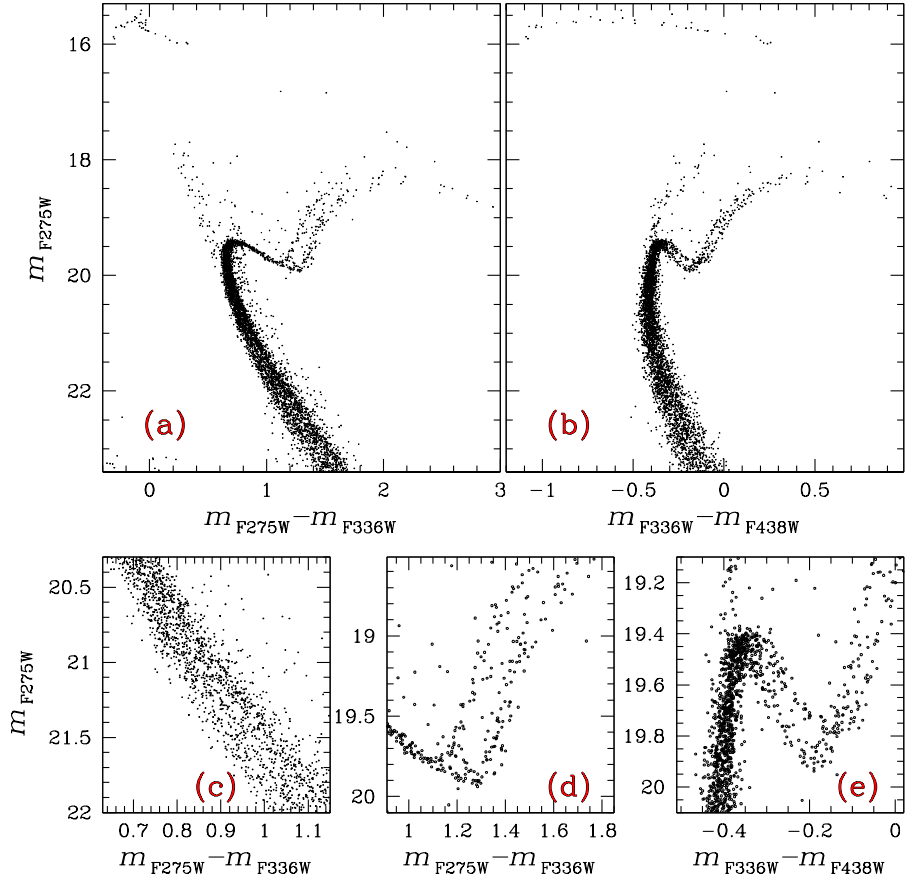}
      \caption{$m_{\rm F275W}$ versus $m_{\rm F275W}-m_{\rm F336W}$ (panel a) 
               and $m_{\rm F275W}$ versus $m_{\rm F336W}-m_{\rm F438W}$ CMD 
               (panel b) of NGC\,288 after differential reddening correction. 
               Panels (c), (d), and (e) are  zoomed-in versions of
               panel (a) and (b)
around the MS, RGB, and SGB, respectively.
}
          \label{CMDs}
   \end{figure}

A visual inspection of the CMDs that we obtain from the data
sets listed in Tab.~\ref{tab:data} indicates that the multiple populations 
along the MS, the RGB, and the SGB are best identified in the $m_{\rm F275W}$ 
versus $m_{\rm F275W}-m_{\rm F336W}$ and the $m_{\rm F275W}$ versus 
$m_{\rm F336W}-m_{\rm F438W}$ CMDs shown in Fig.~\ref{CMDs}.  Panels 
(c) and (d) of the figure show a zoomed-in region around the MS and the RGB, 
and reveal, for the first time, that both the cluster MS and RGB are split 
into two sequences.  Each sequence approximately contains the same number 
of stars. In the following, we will use for the two RGBs and MSs of NGC\,288 
the same nomenclature as previously adopted in our previous works for  the 
cases of 47\,Tuc (Milone et al.\ 2012a), NGC\,6397 (Milone et al.\ 2012c), 
and NGC\,6752 (Milone et al.\ 2013).  In these papers we demonstrated that, 
in the $m_{\rm F275W}$ versus $m_{\rm F275W}-m_{\rm F336W}$ CMD, the 
blue- and the red-RGB stars are the progeny of blue- and red-MS stars, 
respectively. Here, for analogy, we indicate as MSa and RGBa the MS and RGB 
sequence with redder $m_{\rm F275W}-m_{\rm F336W}$ colors, while the bluer 
MS and RGB are named MSb and RGBb, respectively.  The double SGB is 
highlighted in Fig.~\ref{CMDs}e. The two SGBs are well separated in color 
(by $\sim$0.05 mag) in the interval $-0.35<m_{\rm F336W}-m_{\rm F438W}<-0.15$, 
and then merge together at $m_{\rm F336W}-m_{\rm F438W}$$\sim -$0.15, with 
the faint SGB evolving into RGBa. 

\subsection{Population ratio}
\label{sec:cmd}
\begin{figure}[htp!]
   \centering
   \epsscale{.75}
      \plotone{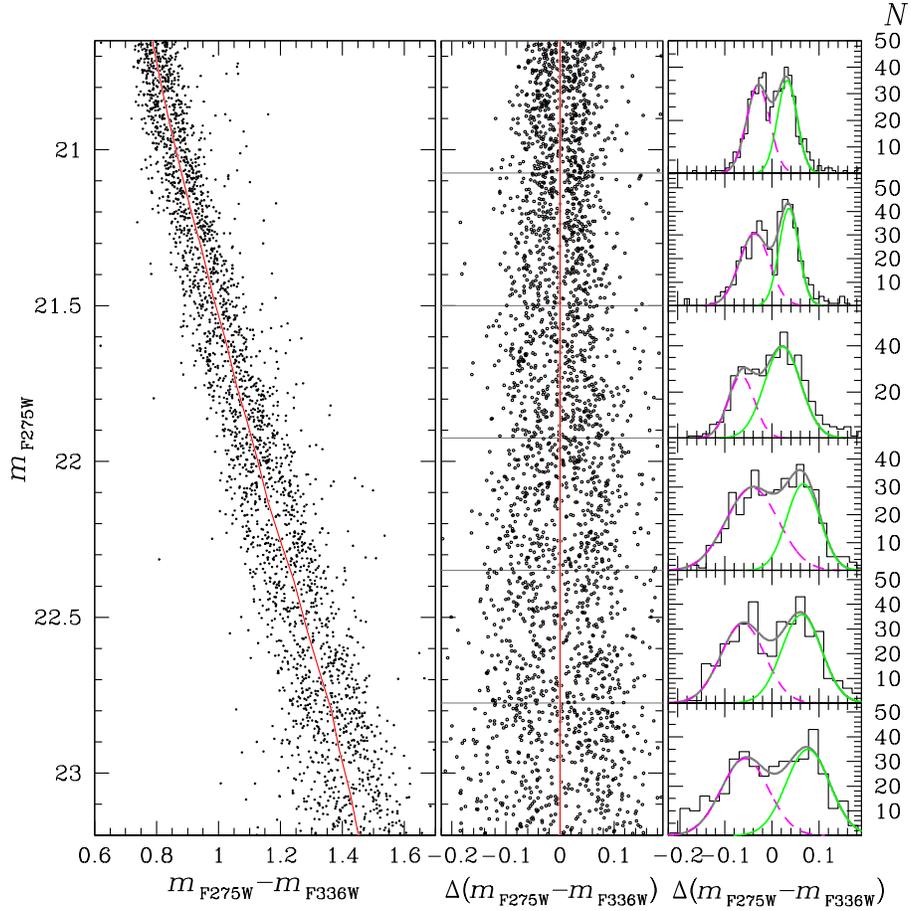}
      \caption{\textit {Left panel}: CMD from Fig.~\ref{CMDs}a zoomed 
               in around the MS region.  The continuous line is a fiducial 
               line for the MS. \textit{Middle panel:} The same CMD, 
               after subtraction of the color of the fiducial line.  
       \textit{Right panels:} Histograms of the $\Delta$ 
               ($m_{\rm F275W}-m_{\rm F336W}$) distribution of the stars, 
               in six magnitude intervals. The continuous gray
               lines show the best fit dual-Gaussian, composed by the 
               sum of the magenta and green Gaussians.}
         \label{msratio}
   \end{figure}

In order to measure the fraction of stars in each MS, we followed the 
procedure illustrated in Fig.~\ref{msratio}, again using techniques
developed in previous studies (e.g.\,Piotto et al.\ 2007, 2012).  
The left panel shows the 
${\it m}_{\rm  F275W}$  vs.\ ${\it m}_{\rm  F275W}-{\it m}_{\rm   F336W}$ 
CMD of Fig.~\ref{CMDs}a, zoomed in around the MS region, in the interval 
20.65$<m_{\rm F275W}<$23.2, where the bimodal distribution is most evident.  
The MS ridge line is marked in red.  To determine it, we started by selecting 
a sample of MS stars by means of a hand-drawn, first-guess ridge line.
We calculated the median color and the median magnitude of MS stars in
bins that were 0.3 magnitude tall.  We then interpolated these median 
points with a spline, and did an iterated sigma-clipping of the 
'verticalized' MS (middle panel).  In order to obtain the  'verticalized' 
MS of the middle panel, we subtracted from each star the color of the 
fiducial line at the same F275W magnitude level, obtaining a  
$\Delta$( $m_{\rm F275W}-m_{\rm F336W}$) value.  The right panels of 
Fig.~\ref{msratio} show the histograms of the distribution of 
$\Delta$ ($m_{\rm F275W}-m_{\rm F336W}$) for six F275W magnitude intervals.

Finally, in each magnitude interval, we fit the histogram with a pair of 
Gaussians, colored green (for the redder peak) and magenta (for the bluer 
peak).  Hereafter, these colors will be consistently used to distinguish 
the MSa and MSb populations and their post-MS progeny.  From the areas 
under the Gaussians we estimate that 54$\pm$3\% of the stars belong to 
the MSa and 46$\pm$3\% to MSb.  The errors were computed from the rms of 
the values obtained for the six intervals.  In the WFC3/UVIS field of 
view, which includes the central part of the cluster with radial distance 
smaller than $\sim$1.2 core radii, the two MSs have almost the same 
number of stars in each magnitude interval, within the statistical 
uncertainties.

\begin{figure}[htp!]
   \centering
   \epsscale{.75}
      \plotone{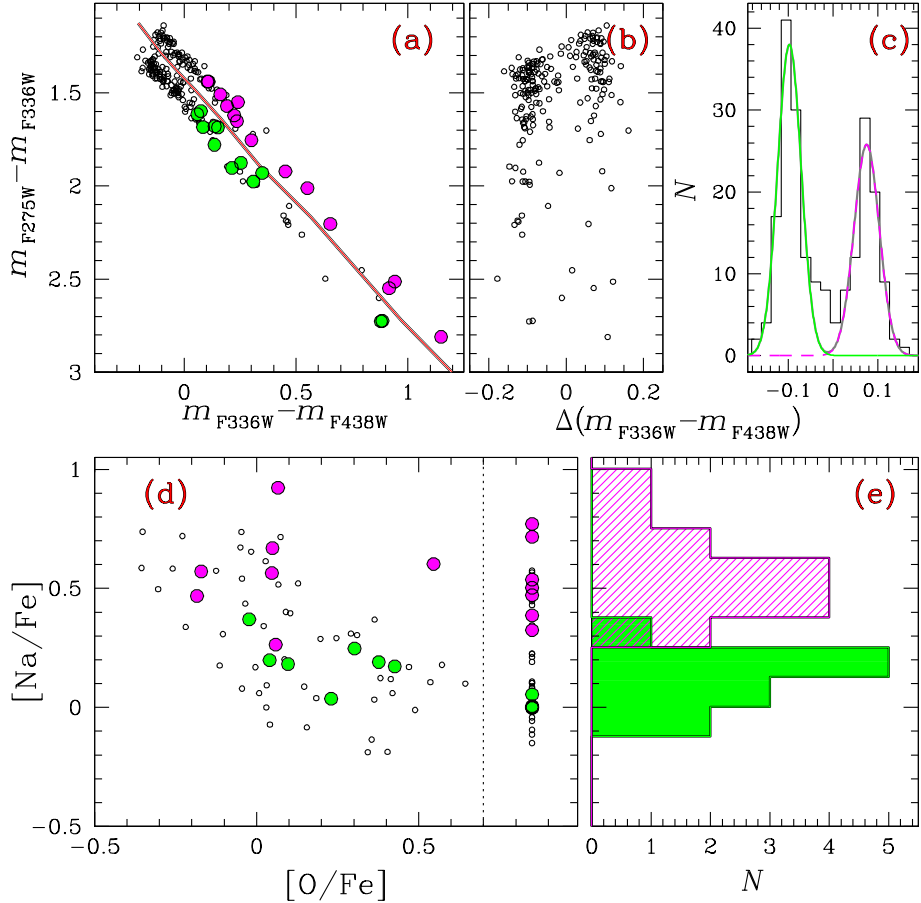}
      \caption{\textit{Panel (a):} $m_{\rm F275W}-m_{\rm F336W}$ 
                  versus $m_{\rm F336W}-m_{\rm F438W}$ two-color diagram 
                  for RGB stars.  The continuous red line is the fiducial 
                  line for the RGB.
               \textit{Panel (b):} Verticalized $m_{\rm F275W}-m_{\rm F336W}$ 
                  versus $\Delta$($m_{\rm F336W}-m_{\rm F438W}$) diagram.
               \textit{Panel (c):} Histogram of the distribution of 
                  $\Delta$($m_{\rm F275W}-m_{\rm F336W}$) for the  stars 
                  shown in the middle panel.  The two components of the 
                  best-fitting dual-Gaussian function are colored green 
                  and magenta.  
               \textit{Panel (d):} Na-O anticorrelation for RGB stars 
                  by Carretta et al.\ (2009). Stars for which only 
                  sodium abundance are available are arbitrarily plotted 
                  at [O/Fe]=0.85 . 
               In panels (a) and (d), RGBa and RGBb stars for which both 
                  spectroscopic and photometric measurements are available 
                  are plotted with green and magenta circles, respectively.
               \textit{Panel (e):} Histogram of the [Na/Fe] distribution 
                  for RGBa (green) and RGBb stars (magenta).  }
         \label{rgbratio}
   \end{figure}

In order to extend the study of stellar populations to the RGB and 
determine the fraction of RGBa and RGBb stars, in Fig.~\ref{rgbratio} we 
show the $m_{\rm F336W}-m_{\rm F438W}$ versus $m_{\rm F275W}-m_{\rm F336W}$ 
two-color diagram, where the RGB of NGC\,288 is clearly split into two 
sequences.  Here, we analyze only RGB stars with $m_{\rm F606W}<17.85$. 
Note that stars are selected on the basis of their F606W magnitude (in order 
to avoid any bias introduced to the strong luminosity difference between RGBa 
and RGBb stars in the ultraviolet filters).  The red line is the hand-drawn
fiducial line for the RGB.  It separates RGBa stars (on the bottom-left 
side) from RGBb stars (on the upper-right side).  We subtracted from the 
$m_{\rm F336W}-m_{\rm F438W}$ color of each star the corresponding color of the
fiducial line, obtaining a $\Delta$($m_{\rm F336W}-m_{\rm F438W}$) index.
The 'verticalized' $m_{\rm F275W}-m_{\rm F336W}$ versus 
$\Delta$($m_{\rm F336W}-m_{\rm F438W}$) diagram is plotted in panel (b) of 
Fig.~\ref{rgbratio}, while panel (c) shows the histogram of the 
$\Delta$($m_{\rm F336W}-m_{\rm F438W}$) distribution. The histogram is 
fitted with the sum of two Gaussians, again colored in green and magenta as
above.  From the area under the Gaussians we calculated that RGBa stars 
include 57$\pm$5\% RGB stars,  with the remaining 43$\pm$5\% stars 
populating the RGBb.  In this case, we simply associated a Poisson error 
to the fraction of stars in each population.  Within one sigma uncertainty, 
these are the same fractions as for the MSa and MSb stars.  From the 
weighted mean of the values obtained from the MS and RGB analysis, we 
obtain that population `a' contains  55$\pm$3\% and population `b' the 
45$\pm$3\% of the total number of stars in the central region analyzed 
in this paper.

Our previous studies of 47\,Tuc and NGC\,6397 have demonstrated that any 
two-color diagram made from the combination of a near-ultraviolet filter 
(such as F225W or F275W), the F336W filter, and a blue filter (such as 
F390W, F435W, or F438W) is particularly efficient at disentangling stellar 
populations with different ligh-element abundances 
(Milone et al.\,2012a,\,c).  These photometric shifts can be interpreted in
the light of spectroscopic observations.
  
Carretta et al.\ (2009) have analyzed GIRAFFE spectra of $\sim$130 stars, 
twenty-five of which are in common with the {\it HST} dataset of this paper.
The spectroscopic targets are represented with large circles in 
Fig.~\ref{rgbratio} and are colored green and magenta according to their 
membership in the RGBa or the RGBb.  The Na-O anticorrelation from 
Carretta and collaborators is reproduced in panel (d), while stars for 
which oxygen-abundance measurements are not available are arbitrarily 
plotted at the flagged value of [O/Fe]=0.85.  
The histogram distributions of [Na/Fe] for RGBa (green  histogram) and 
RGBb stars (magenta histogram) are shown in panel (e).  Similar to what 
is observed in the other GCs studied with a similar approach, we find that 
population `a' stars are Na-poor and O-rich, in contrast to population 
`b' stars, which are depleted in oxygen and enhanced in sodium.

\subsection{A multiwavelength analysis of the double MS}
\label{sec:multiw}
By combining archive and proprietary data, we have access to eleven 
different photometric bands to build CMDs for NGC 288.  
We used the UV and blue photometry displayed in Figs.~\ref{CMDs} 
to select the members of population 'a' and 'b', and then plotted 
their positions in the CMDs obtained with all possible color combinations.
UV photometry  has proven to be essential to separate the two populations, because of its sensitivity to light-element variations (Marino et al.\ 2008).  
On the other hand, optical CMDs are sensitive to He content and allow us to use the color separation of the CMD sequences (MS and RGB) to estimate their average helium difference.  In particular, as shown by  
 Sbordone et al.\ (2011), filters redder than  F435W are marginally 
affected by differences in C N O abundances, while they are sensitive 
to the helium content of the two MSs.

Once we have selected the members of the two populations using the UV 
color-color diagrams, the optical photometry allows us to estimate the 
He content. Helium is extremely difficult to measure by spectroscopy in GC 
stars.  Our procedure below is adapted from that in Milone et al.\, (2013).

\begin{figure}[htp!]
   \centering
   \epsscale{.75}
      \plotone{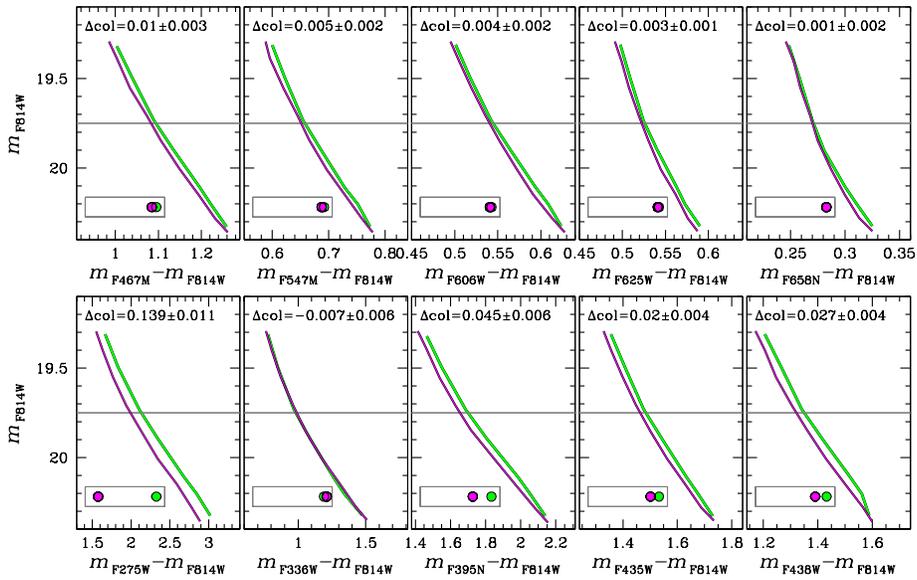}
      \caption{Color-magnitude diagrams for $m_{\rm F814W}$ versus $m_{\rm X}-m_{\rm F814W}$  for MSa (green lines) and MSb (magenta lines) fiducials 
               (X= F275W, F336W, F395N, F435W, F438W, F467M, F547M, 
               F606W, F625W, or F658N). At the top of each panel we 
               give the color distance from the MSa, measured at 
               $m_{\rm  F814W}^{\rm cut}$=19.75 (solid line).  The inset 
               of each CMDs  spans a total color interval of 0.17 mag and 
               shows the relative positions of MSa and MSb represented 
               as green and magenta circles, respectively, at 
               $m_{\rm  F814W}^{\rm cut}$=19.75.  }
          \label{fiducials}
   \end{figure}

Fig.~\ref{fiducials} shows the fiducial ridge lines for the MSa and MSb 
stars in the CMDs constructed with
 ${\it m}_{\rm F814W}$ vs.\ ${\it m}_{\rm X}-{\it m}_{\rm F814W}$ 
(where X= F275W, F336W, F395N, F435W, F438W, F467M, F547M, F606W, F625W, 
or F658N).  A visual inspection reveals that MSa is generally redder than 
MSb, with the only exception of the 
${\it m}_{\rm F814W}$ vs.\ ${\it m}_{\rm F336W}-{\it m}_{\rm F814W}$ 
baseline.  The separation of the two sequences increases for larger color 
baselines in the remaining CMDs, in close analogy with what has been 
observed in the cases of $\omega$\,Cen, NGC\,6397, 47\,Tuc, and NGC\,6752.

\begin{figure}[htp!]
   \centering
   \epsscale{.45}
      \plotone{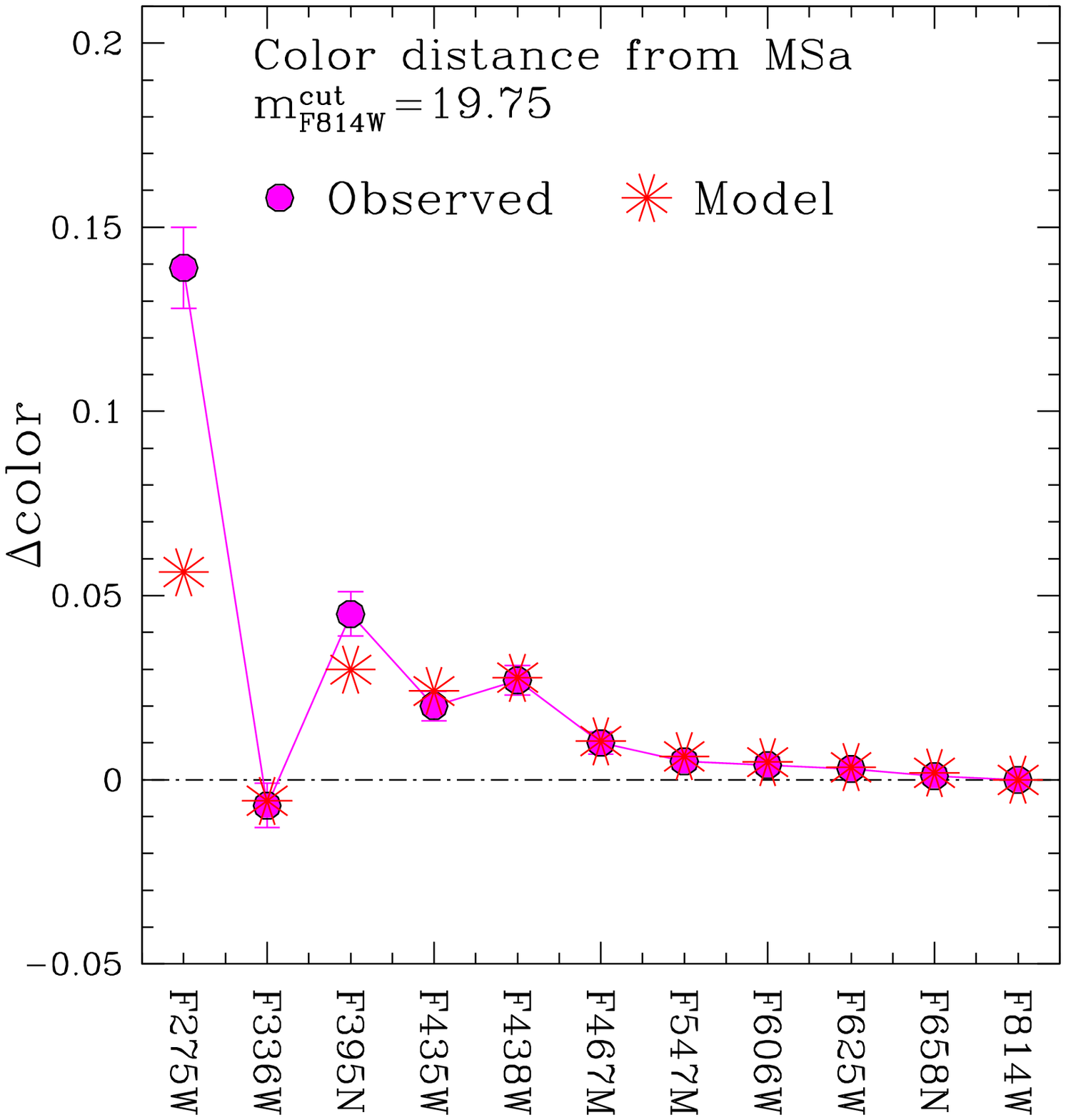}
      \plotone{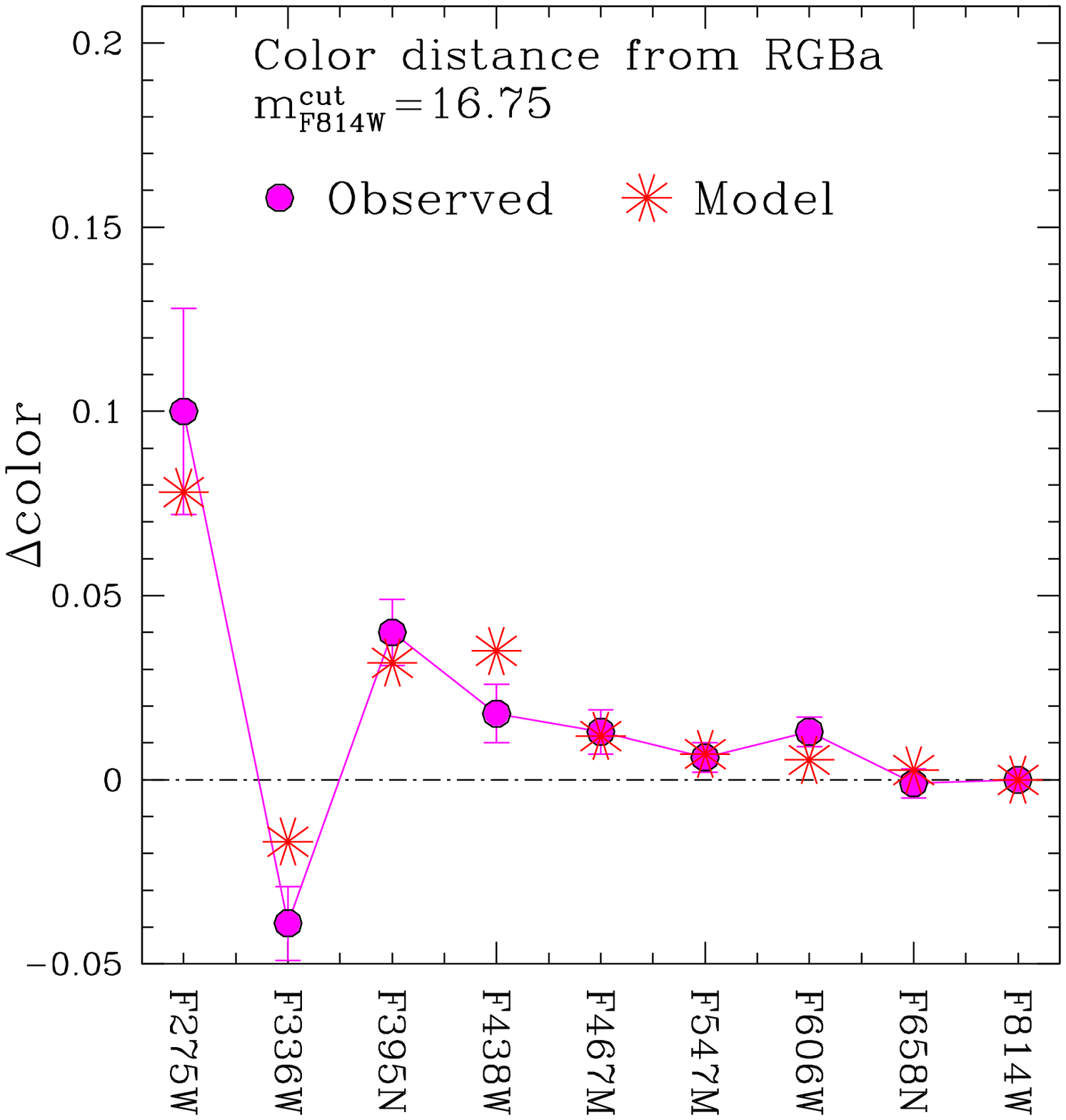}

      \caption{Observed $m_{\rm X}-m_{\rm F814W}$ color separation 
               between MSa and MSb (left panel) and between RGBb and RGBa 
               (right panel) for the available filters (magenta filled circles). 
               Red asterisks indicate the synthetic colors corresponding 
               to the best-fitting models.  The color distances between 
               the MS and RGB fiducials are measured at the reference 
               magnitude $m_{\rm F814W}^{\rm cut}$=19.75, and 
               $m_{\rm F814W}^{\rm cut}$=16.75, respectively.}
          \label{MSdis}
   \end{figure}

Finally, we quantified the MS separation by measuring the color difference 
between MSa and MSb fiducials at a reference magnitude 
$m_{\rm F814W}^{\rm cut}$.  We repeated this procedure for 
$m_{\rm F814W}^{\rm cut}=19.35, 19.55, 19.75, 19.95, 20.15$.  As an 
example, we show the color differences for the case of 
$m_{\rm F814W}^{\rm cut}=19.75$ in Fig.~\ref{MSdis} .

We followed the same procedure as used for the MS to analyze the color 
separation of RGBa and RGBb. Due to the relatively small number of RGB stars, 
we calculated the distance between the two RGB fiducials for the two values 
of  $m_{\rm F814W}^{\rm cut}= 17.25$ and $16.75$.  In close analogy to the 
color behavior of the two MSs, RGBb is typically bluer than the RGBa, with the
 exception of  CMDs based on the $m_{\rm F336W}-m_{\rm F814W}$ color.  In 
the other filters the color distance from the RGBa of the RGBb increases 
with the color baseline. Results are illustrated in the right panel of 
Fig.~\ref{MSdis} for  $m_{\rm F814W}^{\rm cut}= 16.75$.

Fig.~\ref{MSdis} with synthetic photometry predictions.  We used BaSTI 
isochrones (Pietrinferni et al.\ 2004, 2009) to calculate the surface 
temperature ($T_{\rm eff}$) and gravity ($\log g$) at different 
$m_{\rm F814W}$=$m_{\rm F814W}^{\rm cut}$ for two MS populations with 
helium abundances as listed in Table~2.  Table~2 also gives the resulting
($T_{\rm eff}$) and gravity ($\log g$).  In our calculation, we assumed 
$E(B-V)=0.03$ and  $(m-M)_{\rm V}=14.84$  (Harris 1996, 2010 edition). 
We used the average [O/Fe] values for population `a' and population `b' 
stars derived by the measurements in Carretta et al.\ (2009): [O/Fe]=0.2 
and  [O/Fe]=$-$0.0 for first- and second-generation stars, respectively.
Since neither carbon nor nitrogen abundance estimates were available 
for NGC\,288, we  arbitrarily assumed that the MSa has solar N and  C, 
while MSb stars has a carbon depletion of 0.15 dex ([C/Fe]=$-0.15$) and 
enhanced in nitrogen by 0.7 dex ([N/Fe]=$+0.7$).  To avoid the possibility
that the adopted (and uncertain) values of [N/Fe] and [C/Fe] could affect 
our conclusions, we estimate helium only from filters redder than F435W. 
We assumed for the MSa primordial helium content (Y=0.246) and assumed 
for the MSb different helium abundances, with Y ranging from 0.246 
to 0.300 in steps of $\Delta$Y=0.001.

We used the ATLAS12 program (Kurucz 2005, Castelli\,2005, 
Sbordone et al.\ 2007) to account for the adopted chemical composition 
and performed spectral synthesis from $\sim$2,000 \AA\ to $\sim$10,000 \AA\ 
by using the SYNTHE code (Kurucz\,2005).  Synthetic spectra have been 
integrated over the transmission curves of the appropriated filters, and, 
for each value of Y of our grid, we calculated the color difference 
$m_{\rm X}-m_{\rm F814W}$. 

The best fit between models and observations was determined by means 
of chi-square minimization.  The helium difference corresponding to 
the best-fit models are listed in Table~2 for each adopted 
$m_{\rm F814W}^{\rm cut}$ value. From the average mean we obtain that 
population `b' is helium-enhanced by $\Delta$Y=0.013$\pm$0.001, where 
the error is calculated from the agreement of the independent measurements.
Results are shown in Fig.~\ref{MSdis} for the case of 
$m_{\rm F814W}^{\rm cut}=19.75$, and  $m_{\rm F814W}^{\rm cut}=16.75$. 
Models well match the data for the visual filters, while the agreement 
is poorer for the ultraviolet points, as expected since these baselines
are very sensitive to C and N variations, and these abundances are not
constrained by spectroscopy.  
A spectroscopic measure of the C and N for the two populations is clearly needed.

 The C and N abundance differences between population `a' and population `b' stars that we arbitrarily adopted for NGC\,288 are similar to those measured between first and second-generation stars of the GC M\,4 and listed in Tab.~6 by Marino et al.\ (2008, see also Ivans et al.\ 1999, Villanova \& Geisler 2011).
 In order to investigate the impact of our choice of C and N abundances on the inferred helium difference, we repeated the same procedure above by assuming that population `b' stars are nitrogen enhanced by $\Delta$[N/H]=1.0 dex and carbon depleted  by $\Delta$[C/H]=$-$0.5 dex with respect to population `a' stars. In this case, the resulting  $\Delta$Y is consistent with our previous estimate within 0.001 dex, indicating that the conclusions of this paper are not significantly affected by the choice of C and N.

\begin{table}[!htp]
\center
\scriptsize {
\begin{tabular}{ccccccccc}
\hline
\hline
 Sequence & $m_{\rm F814W}^{\rm cut}$ & $T_{\rm EFF}$ (Pop a) & $\log g$ (Pop a)& $T_{\rm EFF}$ (Pop b) & $\log g$ (Pop b)& Y (Pop a) & Y (Pop b) & $\Delta$Y \\
\hline
MS & 19.35 & 6077 & 4.50 & 6100 & 4.49 & 0.248 & 0.262 & 0.014 \\ 
MS & 19.55 & 5966 & 4.54 & 5994 & 4.54 & 0.248 & 0.264 & 0.016 \\
MS & 19.75 & 5840 & 4.58 & 5861 & 4.58 & 0.248 & 0.259 & 0.011 \\
MS & 19.95 & 5701 & 4.62 & 5730 & 4.62 & 0.248 & 0.261 & 0.013 \\
MS & 20.15 & 5558 & 4.65 & 5583 & 4.65 & 0.248 & 0.259 & 0.011 \\ 
\hline
RGB & 16.75 & 5335 & 3.32 & 5347 & 3.31 & 0.248 & 0.265 & 0.017 \\
RGB & 17.25 & 5450 & 3.55 & 5463 & 3.54 & 0.248 & 0.260 & 0.012 \\
\hline
AVERAGE &   &      &      &      &      & 0.248 & 0.261 & 0.013$\pm$0.001 \\
\hline
\end{tabular}
}
\label{tab:elii}
\caption{Stellar parameters of the best-fitting model for population `a' 
         and population `b' stars for different $m_{\rm F814W}^{\rm cut}$
         values.  The helium difference is listed in the last column, 
         while the average $\Delta$Y is given in the list line.}
\end{table}

\section{Summary}
We used multi-band {\it HST} photometry covering a wide range in 
wavelength to study the multiple stellar populations in NGC\,288. 
Once again, UV photometry has proven essential to allow us to separate 
distinct stellar populations.  For the first time, our photometry shows 
that this cluster's MS splits into two branches, and we find that this 
duality is repeated along the SGB and the RGB, similar to what has been 
observed in other GCs.  We calculated theoretical stellar atmospheres 
for main-sequence stars, assuming different chemical composition mixtures, 
and compared the predicted colors through the {\it HST} filters with 
our observed 
colors. 

The observed color differences between the double MS and RGB of NGC\,288 
are consistent with two populations with different helium and light-element 
content.  In particular, population `a', which contains slightly more than 
half of the stars in NGC\,288, corresponds to the first stellar generation 
with primordial He, and O-rich/Na-poor stars, while population `b' is 
made of stars enriched in He by $\Delta Y = 0.013\pm0.001$ (internal error) 
and Na, but depleted in O.  High-precision {\it HST} photometry allows 
us to estimate the He content difference at an accuracy beyond reach 
of spectroscopy.

\begin{acknowledgements}
APM and HJ acknowledge the financial support from the Australian Research 
Council through Discovery Project grant DP120100475.
SC is grateful for financial support from PRIN-INAF 2011 "Multiple 
Populations in Globular Clusters: their role in the Galaxy assembly" 
(PI: E.\,Carretta).
Support for this work has been provided by the IAC (grant 310394), and the Education and Science Ministry of Spain (grants AYA2007-3E3506, and AYA2010-16717). 
GP acknowledges partial support by the Universit\`a degli Studi di Padova CPDA101477 grant.
 JA and AB acknowledge support from STSCI grant GO-12605 
\end{acknowledgements}
\bibliographystyle{aa}

\end{document}